# C-V2X Security Requirements and Procedures: Survey and Research Directions

Vuk Marojevic
Wireless@VT, Bradley Dept. Electrical and Computer Engineering, Virginia Tech
Blacksburg, VA, USA
maroje@vt.edu

*Abstract*—The 3rd Generation Partnership Project (3GPP) defines the first specifications for cellular-based vehicle-to-everything (C-V2X) communications in Release 14. C-V2X extends LTE's device-to-device communication modes by adding two new modes of operation for vehicular systems in coverage and out of coverage of an LTE base station, or eNB. The vehicle-to-vehicle (V2V) communication mode does not rely on the cellular infrastructure and the C-V2X devices employ a distributed and sensing-based semi-persistent scheduling to schedule their packet transmissions. As a promising alternative to dedicated short range communications, the security aspects of C-V2X need to be carefully designed and evaluated to ensure the availability and integrity of the service and data. This paper discusses possible threat scenarios, reviews the 3GPP specifications and finds that, despite the safety-critical nature of C-V2X, only few mechanisms and procedures have be specified to secure the system. We discuss emerging technologies and provide research directions to improve C-V2X system security and reliability and ensure their widespread adoption in civilian and mission-critical communication contexts.

*Keywords—C-V2X, LTE, 3GPP, security.*

## I. INTRODUCTION

Intelligent transportation systems (ITS) will support more efficient vehicular traffic flow, increased vehicular and pedestrian safety, and, eventually, autonomous driving. Wireless communications is fundamental for enabling ITS and recent advances in communications technology and systems enable establishing reliable wireless links and networks among cars, cars and pedestrians, and cars and fixed infrastructure. The success of ITS will be measured in terms of how well it can scale to the ever increasing mobility scenarios and environmental conditions. This poses a stringent need for ultra-reliable and ultra-low latency communications in dense environments, where thousands of cars can be simultaneously present in a given area, moving at different speeds and following different trajectories.

The 3rd Generation Partnership Project (3GPP) specified the long-term evolution (LTE) for providing mobile broadband services in 2008. LTE has been considerably extended in scope since its first release, Release 8 (R8). 3GPP R12 specifies Proximity Services (ProSe) for Device-to-Device (D2D) communications. D2D enables exchange of data over short distances through a direct link between user equipment (UEs). This offers an efficient way to bypass the LTE base station, or eNodeB (eNB), and thus offload the eNB traffic. Apart from content sharing, a D2D UE can act as a relay for another UE with a poor connection to the eNB to extend cellular network coverage. Two modes, Modes 1 and 2, have been defined for *in coverage* and *out of coverage* operation. In Mode 1, the eNBs perform the scheduling of UEs, and each UE needs to be within the serviceable area of an eNB. In Mode 2, distributed scheduling is carried out by the D2D UEs, which do not need to be in the coverage area of the LTE radio access network. One of the operational principle of Modes 1 and 2 is battery life improvement of mobile devices. But, D2D is also an important mode for mission-critical networks since it allows creating ad-hoc networks where there is no cellular infrastructure or where the infrastructure is damaged. Next generation public safety networks will be LTE-based and make use of UEs with D2D communications capabilities.

The constraints and requirements of vehicular communications are quite different than those of static or slowly moving D2D users. Specifically, the high latencies of ProSe are not suitable for vehicular communications, where packet delays or packet losses can have severe consequences. 3GPP R14 therefore extends the ProSe functionality by adding two new modes, Modes 3 and 4, for cellular vehicle-to-everything (C-V2X) connectivity. Mode 3 encompasses vehicle-to-infrastructure (V2I) and vehicle-to-network (V2N), whereas Mode 4 supports vehicle-to-vehicle (V2V) and vehicle-to-pedestrian (V2P) communications [1]. In other words, Mode 3 uses the radio access network, whereas Mode 4 enables UEs to directly talk to one another. These two C-V2X communications modes have been designed to satisfy the latency requirements and accommodate high Doppler spreads and high density of vehicles. They are the vehicular communications equivalents of D2D Modes 1 and 2.

C-V2X has been introduced for safety and non-safety related applications. Safety applications include the dissemination of accidents and sudden breaks. Non-safety applications enable operational and resource efficiencies, among others, by providing relevant information about road status, traffic lights, and so forth. Both types of applications require reliable and timely reception of messages that are confidential and integrity protected. Hence, security is another key requirement for C-V2X and constitutes a challenging research problem because of the stringent resource constraints as well as the dynamics of C-V2X systems and applications. C-V2X messages are short and are regularly broadcast from vehicular UEs whose recipients change. Hence, the traditional signaling and processing procedures to authenticate and authorize transmitters and protect messages are not suitable in this context and alternative solutions need to be sought.

This paper describes some potential threats to C-V2X communications, discusses the proposed security mechanisms in the standards, and derives recommendations and research directions.

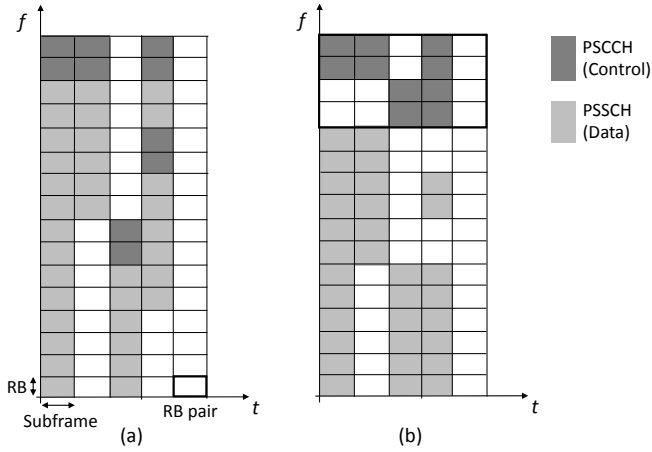

**Fig. 1.** Adjacent (a) and nonadjacent (b) PSCCH and PSSCH.

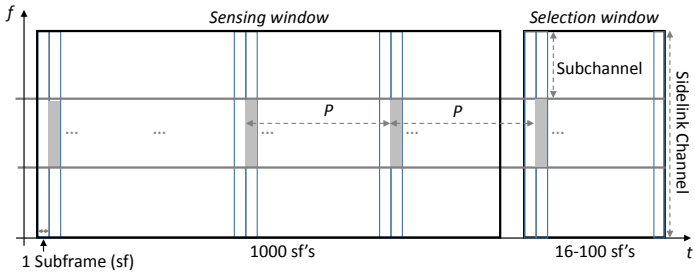

**Fig. 2.** Sensing and selection windows for SPS in C-V2X.

Section II introduces the operational principles of C-V2X. Sections III and IV discuss the need for C-V2X security and privacy, the threats, and the requirements and procedures as standardized by 3GPP. Section V discusses alternative solutions and provides research directions. Section VI concludes the paper.

The scope of this paper is limited to LTE-based V2X, or C-V2X, and the analysis of the security requirements, protocol-specific procedures and research opportunities with focus on the communications layers. Non-protocol specific cyberattacks on vehicles or ITS systems [2] [3] are beyond the scope of this paper. It should be noted that other types of ITS communications standards exist [4] [5], dedicated short-range communications (DSRC) [6] being the most prominent.

## II. C-V2X RESOURCES AND COMMUNICATIONS PROCEDURES

C-V2X uses the sidelink (SL) for sending and receiving messages between a vehicular UE and an eNB (Mode 3) or among vehicular UEs (Mode 4). The former uses the LTE-Uu interface, whereas the latter uses the new PC5 interface, the direct link between two vehicular UEs. The maximum allowed latency ranges between 20 and 100 ms, depending on the application.

### A. Resources

Sidelink resources are shared with the LTE uplink (UL). C-V2X allows channel bandwidths of 10 and 20 MHz. A channel is further divided into subchannels in the frequency domain and subframes in the time domain. Each subchannel consists of several resource blocks (RBs). An RB is defined as 12 subcarriers by 1 slot, or 180 kHz by 0.5 ms to be consistent with the LTE UL or downlink (DL) operation. The number of RBs per subchannel is configurable.

Maintaining LTE's transmission time interval (TTI) of 1 ms, the 1 ms subframe constitutes the time granularity of C-V2X messages and is the basis for message scheduling. As with LTE, RBs are thus scheduled in pairs.

The vehicular UE data is carried over the Physical SL Shared Channel (PSSCH), which is transmitted along with the Physical SL Control Channel (PSCCH) that carries the SL control information (SCI). Two resource configurations are possible: In the adjacent configuration (Fig. 1a), the control and data channels are transmitted in adjacent RBs, occupying one or several subchannels. The SCI is transmitted in the first two RBs in the reserved subframe—four in total—followed by the data. In the nonadjacent configuration, two separate resource pools are configured for the PSCCH and PSSCH (Fig. 1b).

### B. Sidelink Control Information

SCI Format 1 is introduced in 3GPP R14 for C-V2X, whereas SCI Format 0 remains from R12 for D2D. The SCI is transmitted over the PSCCH and carries the information related to the transmission of data over the PSSCH. SCI Format 1 informs the receiving vehicular UEs about the resource reservation interval, the frequency location of initial transmission and retransmission, the time gap between initial transmission and retransmission, and the modulation and coding scheme of the PSSCH data.

### C. Physical Layer

As for the LTE UL, the C-V2X channel access is singe-carrier frequency division multiple access (SC-FDMA). The control information is QPSK modulated; QPSK or 16-QAM is used for the data. Only normal cyclic prefix is supported in C-V2X. The last SC-FDMA symbol in each subframe serves as a guard period.

C-V2X increases the number of subframes carrying the Demodulation Reference Signal (DMRS) to three per RB for the Physical SL Broadcast Channel (PSBCH) and to four for the PSCCH and PSSCH to account for the high Doppler spread at relative speeds of up to 500 km/h [7].

### D. MAC Layer: Scheduling

In Mode 3, the UEs communicate with the infrastructure or network via eNBs, using the LTE Uu interface. All the control and data signaling is between the C-V2X-enabled UE and the eNB, which also handles the scheduling and centrally controls the access to the radio spectrum. In Mode 4, however, scheduling is decentralized and is within the responsibility of the vehicular UEs. More precisely, UEs continuously sense the spectrum and select resources for transmission, employing the semi-persistent scheduling (SPS) [8].

SPS was introduced in earlier LTE releases to support services that require deterministic latency, such as voice. Mode 4 uses it to determine suitable semi-persistent transmission opportunities, i.e. the set of physical subframes and subchannels for regular transmission of messages over a certain time period.

As illustrated in Fig. 2, vehicular UEs continuously sense the radio frequency (RF) spectrum and use the decoded data over one thousand past subframes to select a set of candidate single-subframe resources within the selection window. A candidate single-subframe resource, which is illustrated as the shaded block in the selection window of Fig. 2, consists of the RBs contained in a subframe and one or more contiguous subchannels. The number of required subchannels is a function of the message size.

The SPS scheduling procedure is as follows: The UE initially considers all resources as candidate resources. A two-step exclusion process first excludes those single-subframe resources that have not been monitored. The SPS algorithm then uses the sensing information to discard additional single-subframe resources where the reference signal received power (RSRP) is found to be above the established threshold. If more than 20% of the total resources in the selection window remain, the MAC randomly selects a candidate resource from the 20% of single-subframe candidate resources with the smallest average SL received signal strength indicator (S-RSSI). If fewer the 20% of candidate resources remain in the pool, the threshold is increased by 3 dB and the two-step exclusion process is repeated.

Resources are maintained for a certain number of message transmission intervals. This number is randomly selected in a range that depends on the message rate. Between 5 and 75 transmissions or multiples thereof can be scheduled by the UE with a single SPS invocation [9].

The detailed SPS procedure is specified by 3GPP in [7] [9] and analyzed in [8].

### III. C-V2X Security and Potential Threats

There are different communications security aspects that are desired for ITS. These can generally be classified as [2]:

- *Identification and authenticity* of the user to enable authorized access to services or information as well as authorized provisioning of services or information,
- *Integrity* of messages to ensure that information is accurate and can be trusted,
- *Availability* of the service or information,
- *Confidentiality and privacy* of users, their data and actions from eavesdropping and exploitation, and
- *Non-repudiation and accountability* of the source.

In C-V2X, users broadcast messages and must be authorized to do so. The data and control signaling they broadcast need to be legitimate and accurate. Privacy is important to avoid tracking of users and exploits, among others. Table I outlines some of the C-V2X security threats, which are discussed in continuation under the above categories.

#### A. Identification, Authenticity, and Integrity

The messages that C-V2X UEs exchange are related to information about vehicle speeds, directions, and other actions, such as breaking or accelerating. Such information is encapsulated in basic safety messages (BSMs). BSMs, introduced by the Society of Automotive Engineers (SAE) in the US, support all safety related V2X applications. These messages are regularly broadcast, usually at a rate of 10 Hz. Event-triggered messages inform about sudden or unexpected events, such as accidents. The messages that are transmitted from one vehicle will thus affect the operation of other vehicles in the area. If a UE transmits false information, other UEs receiving it may trigger actions, which, instead of optimizing the vehicular traffic flow on roads and highways, may generate chaos. For instance, if a UE warns of an accident that does not exist, the approaching vehicles may slow down creating congestion. When this happens on recurring basis, the level of thrust will decrease and future messages may be ignored, defeating the purpose of C-V2X as a technology to increase traffic safety, resource efficiency, and system performance. It is therefore paramount that messages can be trusted. Since they are transmitted over the air and heard by many, source and message authenticity mechanisms need to be enforced.

#### B. Availability

Capacity is a key metric in wireless communications and C-V2X. Different techniques exist to increase capacity, but irrespective of these, adding more users adds interference and stress to the network because of the physically limited and shared resources. If dummy users create service requests or participate in broadcasting dummy messages, the RF spectrum will become more congested and the C-V2X service less reliable, compromising the availability of information.

As shown in other communications contexts, including LTE, radiating in RF bands where C-V2X services are provided, can make the proper decoding of messages challenging, if not impossible. Mechanisms that allow recovering corrupted or lost messages, usually by means of strong coding and retransmission, add to the RF congestion, increase latency and power consumption, among others.

#### C. Confidentiality and Privacy

The identifies, position, actions and trajectories of UEs need to be confidential to avoid tracking of vehicles. If messages from UEs are replayed at different times or at different locations, RF congestion is the immediate effect, but also confusion if two messages from the same source are received with inconsistent information. Jamming the regular transmissions from a single vehicular UE is also possible, even without message decoding because of the structured SPS and message rates. In dynamic vehicular environments, confusion about a vehicle location, speed, etc. can be created through delayed message replay without modifying the message content.

TABLE I – SOME ENVISIONED C-V2X SECURITY THREATS AND SCENARIOS.

| Threat | Scenario | Key |
|---|---|---|
| Fake nodes | Fake UEs, UE-road side units (RSUs), or eNB-RSUs can be deployed and this is a legitimate threat occurring today with 4G LTE [11] [12]. Such nodes can compromise UE privacy or provide false control information or data. | *A.* *C.* |
| False information | Rogue UEs can provide false information to deceive other UEs and trigger certain actions. | *A.* |
| Fake certificates | Rogue UEs can send many fake authentication certificates to drain the computing resources of UEs trying to verify the sender [13]. | *A.* *B.* |
| RF congestion | Illegitimate UEs that transmit frequent messages will create more congestion in the spectrum, leading to more interference and lower system performance. | *B.* |
| Jamming | Jammers create direct interference to transmitted signals to impede their proper demodulation. Different types of jammers exist and can cause system degradation at different levels [10]. | *B.* |
| RF replay | UEs that receive and decode messages from other UEs can replay them. This can cause congestion, (reactive) jamming, or confusion about inconsistent message content. One can imagine a network of distributed UEs that replay messages and cause congestion, confusion, or both in different areas. | *C.* |
| Malfunctioning UE | A UE listens to broadcasted messages as well as senses the RF environment to broadcast. If (1) time or frequency synchronization is not properly established, (2) RF sensing is not reliable due to equipment failure, or (3) the SPS or congestion control mechanisms fails, significant interference can be caused because of the time and frequency sensitive nature of orthogonal frequency division multiple access. | *D.* |

*D. Non-Repudiation and Accountability*

Malfunctioning UEs can significantly compromise SC-FDMA system performance. 3GPP establishes mechanisms when non-C-V2X UEs are out of sync. For example, a UE is not allowed to transmit if it stops receiving the LTE time advancement commands from the eNB. Similar mechanisms are needed for C-V2X and need to be enforced. In C-V2X, UEs rely mostly on GPS as the synchronization source, but can use infrastructure nodes, where available, or other UEs. If no external source exists, depending on the type and quality of oscillators, frequency and timing drifts will occur that add up over time and can cause system malfunctioning and significant interference. Harmful nonsynchronous transmission needs to be identified and the corresponding UEs accounted for it.

IV. 3GPP C-V2X SECURITY REQUIREMENTS AND PROCEDURES

This section analyzes the 3GPP security requirements and procedures for C-V2X.

*A. Authentication Requirement*

3GPP states that *"V2X network entities shall be able to authenticate the source of the received data communications, of data between V2X network entities shall be confidentiality and integrity protected and protected from replays"* [14]. Only authorized entities should be able to transmit data and only messages whose data origin has been verified should be processed by the UE. Since it is not possible to prevent RF transmissions from unauthorized users, including standard-compliant transmissions that would appear regular and legitimate, the recipients should be able to identify the sender and verify if it was authorized to transmit the data.

3GPP thus recognizes the fundamental need for user authentication, which is essential for trust. If the data comes from an unauthorized or unknown transmitter, it may contain unreliable information and should be handled with caution. The specifications, however, only outlines the requirement, not the mechanisms and different manufacturers may feel different about the need for strong authentication and may have other business priorities.

Traditional LTE authentication mechanisms can be employed and enforced in Mode 3. But, an operator may not be present in Mode 4 and proper authentication has to rely on the chip manufacturers and exhaustive testing. Standardization is important for interoperability and for setting appropriate authentication standards for all participating UEs as one weak vehicular UE may compromise the security of the entire system.

*B. Privacy Requirement*

3GPP states that UE identity should not be long-term trackable or identifiable from its transmissions over the PC5 interface. Although UE authenticity needs to be verifiable, as discussed above, this should be used only for the purpose of verifying the legitimacy of the sender (and the message). This requirement applies to UEs and non-V2X entities and extends to operators and third parties.

In order to achieve the above, the permanent identities of UEs need to be properly protected and their exposure minimized using pseudonymity. Fake eNBs can force legitimate UEs to share their international mobile subscriber identity (IMSI) and location information [11] and equivalent attacks could affect vehicular UEs.

3GPP also requires protection from eavesdropping of the application layer UE identity, which is transmitted as part of the V2X message.

Whereas these requirements are appropriate, it is less clear how to effectively achieve this in an ad-hoc network (C-V2X Mode 4), as opposed to the infrastructure-supported solution (Mode 3). More precisely, 3GPP does not impose any privacy mechanisms for the PC5 SL and leaves this to the regional regulators and operators. It suggests changing and randomizing the source layer 2 ID and source IP address in coordination with the application layer ID changes. There is no additional protection for the Uu SL beside what LTE provides.

*C. Maintenance Requirement*

Over-the-air (OTA) mechanisms shall be used to secure the transfer of the configuration data and maintain up-to-date security mechanisms. This data will typically be stored in the universal integrated circuit card (UICC).

Whereas security-related updates are critical, OTA updates need special handling. The challenge here is having regular access to trusted sources, especially when roaming or when typically not depending on a central operator (Mode 4). Another challenge is maintaining interoperability among vehicular UEs from different vendors/operators using the highest security standards.

*D. Security Procedures*

The LTE security measures should be used for the LTE-Uu interface, that is, the interface between the UE and the eNB (Mode 3). The network provider decides whether to use ciphering or not. The 3GPP R14 specifications for C-V2X [14] explicitly state that *no security is applied for the PC5 one-to-many communication by setting the fields related to group security to 0*. This means that the basic safety and event-triggered messages that are exchanged among UEs have no standard security mechanisms in place.

Application layer security is outside the scope of 3GPP, but it is suggested that periodically refreshed credentials be used to avoid UE tracking, among others.

Release 15 [15] describes the architectural enhancements for V2X services and provides more details about authorization over both reference points, PC5 and LTE-Uu. While this is indispensable, this alone cannot resolve the concerns identified in Table I, such as how to deal with excessive RF congestion.

V. RESEARCH DIRECTIONS

C-V2X Mode 4 is an ad-hoc communication mode using the PC5 air interface between UEs. In this mode, the peers are not known a priory and change dynamically because of the different trajectories of vehicles. Since infrastructure is not involved in this mode, standard LTE security mechanisms cannot be applied. 3GPP does not address the establishment of ad-hoc security associations in R14.

Research is needed to identify the key vulnerabilities of a new communications system and propose and demonstrate mechanisms to harden it. Research in the following areas can be leveraged to increase the security of C-V2X:

- **Physical layer techniques:** Use and leverage the diversity of the channel as well as interference mitigation techniques [16]. Embed signatures or other unique identifiers in messages to identify legitimate UEs and legitimate messages.
- **Cross-layer techniques:** Revise scheduling, congestion control and other C-V2X procedures to make them more robust and aware of potential threats. For example, schedule messages for the purpose of enhancing distributed security awareness.

- **Edge and Cloud computing:** Use the processing power of the Cloud to assist in determining fake transmitters and leverage edge computing resources to reduce latency and network congestion [17].
- **Network-aided procedures:** Have the network periodically notify UEs of legitimate UEs in the area using C-V2X or parallel communications protocols, such as regular LTE.
- **Radio environment map (REM):** Have UEs built their REMs, where processed RF spectrum activity and the information exchanged among UEs is stored and contrasted to identify anomalies.
- **Spectrum access system (SAS):** The vehicular UEs and roadside units can regularly provide sensing information to a central SAS for RF anomaly detection.
- **Machine learning (ML):** ML can be effectively employed to analyze huge amounts of data and classify it. When either the normal behavior is known or the attacks, or both, ML tools can be trained to distinguish between normal and abnormal activity.
- **Multiple sensor information processing:** Use information from multiple sensors (cameras, radar, lidar, etc.) to validate C-V2X messages, weigh decisions and dynamically update node and information thrust metrics. Check message content for consistency using physical attributes, past messages, environmental information, etc.

One of the main challenges is the real-time nature of the data and the limited resources (time, RF bandwidth, and processing power) that are available. Moreover, the interference or attack may change. Therefore, one solution will not fit all cases and a single approach may result ineffective. Instead, several techniques should be combined and used in concert. For example, the system could do some local processing and combine it with data processed at the Edge or Cloud and data provided by the network. Research has shown that combining multiple sources of data can help identifying and overcoming a threat. An example of this is jointly processing several of the widely used performance measurement counters and key performance indicators in cellular networks to identify a specific source of interference [18].

The following tools can be leveraged for accelerating research, testing, and standardization:
- Software-defined radios and open-source software,
- Laboratory and field experiments, providing data for researchers to use for evaluating new approaches for secure C-V2X systems,
- Crowdsourcing, allowing distributed development and testing as well as contributions to specifications,
- Specifications with more dynamic contributions,
- Regular firmware and software updates and secure software downloads and installations at a rate comparable to software updates of operating systems,
- Use of virtualization technology to enable dynamic network reconfiguration and security as a service.

## VI. CONCLUSIONS

This paper presents the C-V2X communications protocol and analyzes some of its security features and requirements. 3GPP has outlined the security mechanisms, but leaves it in the hands of the developers to implement the proper features. It states: *"All V2X services do not have the same security requirements and hence may not require the use of all the described features. It is up to the deployment of the feature to ensure that all the appropriate security aspects are addressed."*

When compared to D2D, it is more challenging to establish security in C-V2X as it requires very low latency and high reliability in a highly dynamic environment. C-V2X, on the other hand, will typically be less energy and computing resource limited, which means it can collect, store and process more data than handheld D2D UEs.

Ongoing standardization efforts are key for the establishment of reliable communications systems. Security mechanisms need to be incorporated into the C-V2X framework and need to be enforceable. This adds overhead to the resource constrained service and this overhead needs to be accounted for. However, when security is compromised, the system overhead for fixing the issue can become orders of magnitude worse and the immediate and long-term consequences significant. It takes time to revise a standard and redesign a system. If security features are not properly standardized and enabled by design, security patches may not be compatible across platforms and not effective if not implemented by all.

Being a cellular technology based on 4G LTE provides a good foundation because it inherits its security mechanisms and can leverage the abundant and ongoing research in this field. However, 4G LTE has its known vulnerabilities and more research is needed if C-V2X is to become a widely deployed and widely relied on ITS technology. 3GPP R15 introduces 5G New Radio, which will add more security features to make next generation C-V2X more secure. The 5G security architecture and procedures, as specified in [19][20], do not address V2X or D2D out-of-coverage mode of operation yet. Important changes are expected in future revisions, but need continuous research support.